\documentclass[a4paper,11pt]{article}

\usepackage{amsmath,amssymb}
\usepackage{ascmac}
\usepackage{graphicx}
\usepackage[bookmarks=true,bookmarksnumbered=true,bookmarkstype=toc]{hyperref}
\hypersetup{
pdftitle={Gauged Linear Sigma Model with F-term for A-type ALE Space},
pdfauthor={Tetsuji KIMURA and Masaya YATA},
colorlinks={true},
linkcolor={black},
urlcolor={black},
filecolor={black},
citecolor={black}
}
\makeatletter

 \@addtoreset{equation}{section}
\makeatother

\newcounter{Enumerate}

\DeclareFontFamily{U}{rsf}{}
\DeclareFontShape{U}{rsf}{m}{n}{
  <5> <6> rsfs5 <7> <8> <9> rsfs7 <10-> rsfs10}{}
\DeclareMathAlphabet\Scr{U}{rsf}{m}{n}

\usepackage[mathscr]{eucal}

\newcommand{\del}{\partial}

\newcommand{\mfk}{\mathfrak}
\newcommand{\LS}{\ \ \ \ \ \ \ \ \ \ }
\newcommand{\ls}{\ \ \ \ \ }
\newcommand{\wt}{\widetilde}

\newcommand{\ol}{\overline}

\newcommand{\bsubeq}{\begin{subequations}}
\newcommand{\esubeq}{\end{subequations}}

\newcommand{\noi}{\noindent}

\newcommand{\nn}{\nonumber}
\newcommand{\N}{\mathcal{N}}
\renewcommand{\a}{\mfk{a}}
\renewcommand{\d}{{\rm d}}
\newcommand{\e}{{\rm e}}
\renewcommand{\i}{{\rm i}}
\renewcommand{\l}{\ell}

\newcommand{\slb}{\scalebox}
\def\+{{+\!\!\!+}} 

\begin{document}
\allowdisplaybreaks{

\thispagestyle{empty}


\begin{flushright}
KEK-TH-1706,
RUP-14-5
\end{flushright}

\vspace{30mm}

\noi
\slb{2.3}{Gauged Linear Sigma Model with F-term}

\vspace{5mm}

\noi
\slb{2.3}{for A-type ALE Space}

\vspace{15mm}

\slb{1.2}{Tetsuji {\sc Kimura}$^{\,a}$}\footnote{Address since April 2014: Department of Physics, Tokyo Institute of Technology, Tokyo 152-8551, JAPAN. E-mail: {\tt tetsuji \_at\_ th.phys.titech.ac.jp}} \ and \ \slb{1.2}{Masaya {\sc Yata}$^{\,b}$}

\slb{.85}{\renewcommand{\arraystretch}{1.2}
\begin{tabular}{rl}
$a$ & {\sl
Department of Physics \& Research Center for Mathematical Physics,}
\\
&{\sl 
Rikkyo University} 
\\
& {\sl 
Tokyo 171-8501, JAPAN
}
\\
& {\tt tetsuji \_at\_ rikkyo.ac.jp}
\end{tabular}
}

\slb{.85}{\renewcommand{\arraystretch}{1.2}
\begin{tabular}{rl}
$b$ & {\sl
KEK Theory Center, Institute of Particle and Nuclear Studies,}
\\
& {\sl 
High Energy Accelerator Research Organization (KEK)
}
\\
& {\sl 
Tsukuba, Ibaraki 305-0801, JAPAN}
\\
& {\tt yata \_at\_ post.kek.jp}
\end{tabular}
}

\vspace{15mm}


\noindent
\slb{1.1}{\sc Abstract}:
\begin{center}
\slb{.95}{
\begin{minipage}{.93\textwidth}
We construct yet another ${\mathcal N}=(4,4)$ gauged linear sigma model for the $A_N$-type ALE space.
In our construction the toric data of the ALE space are manifest.
Due to the $SU(2)_R$ symmetry, the F-term is automatically determined.
The toric data, which govern the K\"{a}hler structures of the ALE space, are embedded into $U(1)$ charges of charged hypermultiplets.
The F-term is also inevitable to determine the complex structures of the ALE space.
In the IR limit, we obtain the K\"{a}hler potential of the $A_N$-type ALE space.
We also find the origin of the ${\mathbb Z}_{N+1}$ orbifold symmetry in the singular limit of the $A_N$-type ALE space.
In a special case, we reproduce an explicit form of the K\"{a}hler potential of the $A_1$-type ALE space, i.e., the Eguchi-Hanson space.
\end{minipage}
}
\end{center}

\newpage
\section{Introduction}

A gauged linear sigma model (GLSM) is the UV completion of a nonlinear sigma model (NLSM) in the IR regime \cite{Witten:1993yc}.
This is quite a powerful model to investigate vacua and topological aspects of string theory.
If one controls the Fayet-Iliopoulos (FI) parameters, one finds various non-trivial phases.
Some of them provide NLSMs associated with the geometrical aspects of the spacetime in which a string propagates.
The other phases describe conformal field theories (CFT) associated with the topological invariants of string theory.
Indeed they are deeply related to each other.
For instance, a NLSM for a Calabi-Yau variety corresponds to a CFT given by a suitable Landau-Ginzburg (LG) superpotential.
This phenomenon is called the CY/LG correspondence \cite{Vafa:1988uu}.
Applying the Strominger-Yau-Zaslow conjecture \cite{Strominger:1996it} to toric varieties,
one can study various mirror pairs of Calabi-Yau varieties in the framework of the toric GLSM \cite{Hori:2000kt, Hori:2000ck}.
In addition,
the toric GLSM also plays a central role in the analysis of AdS/CFT correspondence \cite{Maldacena:1997re} even in less supersymmetric systems \cite{Martelli:2004wu}. 
The gauge theory duals of the gravity theories on toric varieties have been analyzed in terms of quiver gauge theories \cite{Hanany:1996ie, Douglas:1996sw, Johnson:1996py}, where the toric GLSM is one of the most important tools.

However, one encounters a serious problem when
one investigates physics of string theory beyond the topological aspects.
For instance, it is quite unclear that the perspective of differential geometry of toric varieties are correctly described
(see, for instance, the discussions in \cite{Martelli:2004wu, Oota:2005mr}).
In order to make the problem clear,
let us consider the $A_1$-type ALE space, as a typical example.
In the framework of the $\N=(2,2)$ supersymmetry following \cite{Hori:2003ic},
the GLSM for the $A_1$-type ALE space is given by 
a set of chiral superfields $\{ A_1, A_2, A_3 \}$ whose $U(1)$ charges are $\{ +1, -2, + 1\}$.
If we consider the $A_1$-type ALE space in the viewpoint of algebraic geometry, 
the $\N=(2,2)$ framework is sufficient.
This is the main reason that the $\N=(2,2)$ toric GLSM are often utilized in the analysis of (topological feature of) string theory.

However, the $\N=(2,2)$ supersymmetry is too weak to study it in the viewpoint of differential geometry.
The reason is that
the $\N=(2,2)$ supersymmetry does not restrict F-term, i.e.,
any functions given by superfields are allowed, as far as they are gauge invariant.
This implies that 
we cannot derive the correct NLSM whose target space geometry is the $A_1$-type ALE space in the IR limit.
The toric data are embedded only into the D-term of the GLSM.
The D-term controls the size, or the K\"{a}hler structure, of the target space.
On the other hand, the F-term governs the shape, or the complex structure, of the geometry.  
Thus the most important task to construct the correct GLSM for the ALE space
is to find the correct F-term. 

We recall that the $A_1$-type ALE space is a hyper-K\"{a}hler space.
Since the target space geometry is a hyper-K\"{a}hler,
the $\N=(2,2)$ supersymmetry is extended to $\N=(4,4)$.
Once the $\N=(4,4)$ supersymmetry is involved into the system,
the F-term is automatically generated in order to preserve the $SU(2)_R$ symmetry.
This extension is also applicable to the GLSM for other A-type ALE spaces introduced in \cite{Hori:2003ic}.

In this paper
we construct $\N=(4,4)$ GLSMs for generic A-type ALE spaces in a systematic way.
In our construction the toric data of the ALE space are manifest.
Due to the $SU(2)_R$ symmetry generated by the $\N=(4,4)$ supersymmetry,
the form of the F-term is automatically determined.
In the IR limit of the GLSM,
we can obtain the K\"{a}hler potential of the ALE space, if the FI parameters vanish.
This formulation will be quite useful to analyze string theory on the ALE spaces, aspects of five-branes, and so forth. 
We emphasize that our Lagrangian is yet another $\N=(4,4)$ GLSM different from the well-known formulation \cite{Douglas:1996sw, Johnson:1996py, Giveon:1998sr, Belhaj:2000ai, Tong:2002rq, Harvey:2005ab, Okuyama:2005gx}.

\vspace{3mm}

The structure of this paper is as follows:
In section \ref{GLSMF-A1},
we construct the GLSM for the $A_1$-type ALE space.
This is the simplest ALE space.
We first mention the $\N=(2,2)$ system associated with the toric data.
This is immediately extended to the $\N=(4,4)$ system.
Introducing additional matter fields and a vector multiplet,
we exhibit the $\N=(4,4)$ GLSM with F-term.
Next, we analyze the dynamics in the IR limit.
Due to the existence of the field equations from the F-term,
a discrete symmetry is emerged.
Finally, integrating out all the vector multiplets,
we obtain the NLSM given by the K\"{a}hler potential of the $A_1$-type ALE space.
We emphasize that 
the discrete symmetry of the solution in the gauge theory 
becomes the orbifold symmetry of the geometry in the singular limit.
In section \ref{GLSMF-AN},
we study the GLSM for the $A_N$-type ALE space.
This is a natural extension of the GLSM for the $A_1$-type ALE space.
Since it is difficult to solve the equations of motion for the vector multiplets in the presence of FI parameters, 
we focus only on the case that all of them vanish.
In this case we obtain the NLSM for the singular limit of the $A_N$-type ALE space.
Section \ref{Summary} is devoted to summary and discussions.
In appendix \ref{GLSMF-A2},
we discuss the $\N=(4,4)$ GLSM for the $A_2$-type ALE space.
In this model we explicitly construct the K\"{a}hler potential of the geometry where the two singularities are blown up.

\section{$A_1$-type ALE space: Eguchi-Hanson}
\label{GLSMF-A1}

In this section 
we study the $\N=(4,4)$ GLSM for the $A_1$-type ALE space, i.e., the Eguchi-Hanson space.
This geometry is described as the ${\mathbb C}^2/{\mathbb Z}_2$ orbifold,
or the ${\cal O}(-2)$ bundle over ${\mathbb C}{\bf P}^1$ if the singularity is blown up.
We start from the $\N=(2,2)$ GLSM for the $A_1$-type ALE space 
associated with the toric data \cite{Hori:2003ic}.
The theory contains an abelian vector superfield $V_1$ and chiral superfields $\{ A_1, A_2, A_3 \}$
whose $U(1)$ charges are $\{ +1, -2, +1 \}$, as mentioned before.
This charge assignment is associated with the condition that the first Chern class vanishes.
If one considers the $A_1$-type ALE space in the viewpoint of algebraic geometry, 
the $\N=(2,2)$ framework is sufficient.
However, if one studies it in the viewpoint of differential geometry, 
the $\N=(2,2)$ field contents should be extended to the $\N=(4,4)$ field contents exhibited in Table \ref{table:A1}:

\begin{table}[h]
\begin{center}
\slb{.9}{${\renewcommand{\arraystretch}{1.05}
\begin{array}{c|ccc}
\text{$A_{1}$-type} & (A_1, B_1) & (A_2, B_2) & (A_3, B_3) 
\\ \hline\hline
(V_1, \Phi_1) & (+1,-1) & (-2,+2) & (+1,-1) 
\\ 
(\wt{V}, \wt{\Phi}) & 0 & (-\alpha, \alpha) & 0
\end{array}
}$} 
\end{center}
\caption{Field contents of $\N=(4,4)$ GLSM for $A_1$-type ALE space.}
\label{table:A1}
\end{table}

The constituents of the $\N=(4,4)$ GLSM for the $A_1$-type ALE space are 
three charged hypermultiplets and two vector multiplets.
Let us explain them in details:
The three charged hypermultiplets in the $\N=(4,4)$ system are given by 
sets of the $\N=(2,2)$ chiral superfields $\{ A_i, B_i \}$.
They are doublets under the $SU(2)_R$ symmetry.
The $\N=(4,4)$ vector multiplet contains
an $\N=(2,2)$ vector superfield $V_1$ and a neutral chiral superfield $\Phi_1$.
We also introduce an additional $\N=(4,4)$ vector multiplet $\{ \wt{V}, \wt{\Phi} \}$ in order to remove redundant degrees of freedom.
The components in Table \ref{table:A1} denote the $U(1)$ charges of respective gauge symmetries, 
where $\alpha$ is arbitrary except for zero.
Due to the $\N=(4,4)$ supersymmetry, the Lagrangian of the gauge theory with F-term is completely determined as follows\footnote{The conventions of the $\N=(2,2)$ supersymmetry in two dimensions are subject to the recent works \cite{Kimura:2013fda, Kimura:2013zva, Kimura:2013khz}.}: 
\begin{align}
\Scr{L}
\ &= \
\int \d^4 \theta \, \Big\{
\frac{1}{e_1^2} \Big( - |\Sigma_1|^2 + |\Phi_1|^2 \Big)
+ \frac{1}{\wt{e}^2} \Big( - |\wt{\Sigma}|^2 + |\wt{\Phi}|^2 \Big)
\Big\}
\nn \\
\ & \ \ \ \ 
+ \int \d^4 \theta \, \Big\{
|A_1|^2 \, \e^{2 V_1}
+ |A_2|^2 \, \e^{-4 V_1 - 2 \alpha \wt{V}}
+ |A_3|^2 \, \e^{2 V_1}
\Big\}
\nn \\
\ & \ \ \ \ 
+ \int \d^4 \theta \, \Big\{
|B_1|^2 \, \e^{-2 V_1}
+ |B_2|^2 \, \e^{4 V_1 + 2 \alpha \wt{V}}
+ |B_3|^2 \, \e^{-2 V_1}
\Big\}
\nn \\
\ & \ \ \ \ 
+ \Big\{
\sqrt{2} \int \d^2 \theta \, \Big[
\Phi_1 \Big( - A_1 B_1 + 2 A_2 B_2 - A_3 B_3 - s^1 \Big)
+ \wt{\Phi} \Big( \alpha A_2 B_2 - \wt{s} \Big)
+ \text{(h.c.)}
\Big]
\Big\}
\nn \\
\ & \ \ \ \ 
+ \Big\{
\sqrt{2} \int \d^2 \wt{\theta} \, \big( - t^1 \Sigma_1 - \wt{t} \, \wt{\Sigma} \big)
+ \text{(h.c.)}
\Big\}
\, . \label{GLSMF44-A1}
\end{align}
Here $e_1$ and $\wt{e}$ are the gauge coupling constants of mass dimension one.
The kinetic terms and the twisted F-term of the vector superfields are given by $\Sigma = \frac{1}{\sqrt{2}} \ol{D}{}_+ D_- V$.
We introduced the complexified FI parameter 
$t^1 = \frac{1}{\sqrt{2}} (t^1_1 + \i t^1_2)$ associated with $\Sigma_1$.
This is nothing but the blown-up parameter of the singularity.
It is possible to introduce another complex parameter $s^1$ associated with $\Phi_1$.
The pair $\{ t^1, s^1 \}$ becomes a doublet under the $SU(2)_R$ symmetry.
For simplicity, however, we set $s^1$ to zero in this work.
We also set the FI parameters $\{ \wt{t} , \wt{s} \}$ associated with $\wt{\Sigma}$ and $\wt{\Phi}$ to zero.
We notice that the coefficients in the F-term are determined by the $SU(2)_R$ symmetry.


Consider the IR dynamics of the GLSM.
Since the gauge coupling constants go to infinity in the IR limit,
the vector multiplets become auxiliary fields.
In order to integrate them out from the system, 
we evaluate the equations of motion.
The equations of motion for $V_1$ and $\wt{V}$ are
\bsubeq \label{EOM-V-A1}
\begin{align}
\a^2
\ &= \ 
|A_1|^2 \, \e^{2 V_1} 
- |B_1|^2 \, \e^{-2 V_1}
+ |A_3|^2 \, \e^{2 V_1}
- |B_3|^2 \, \e^{- 2 V_1}
\nn \\
\ & \ \ \ \ 
- 2 |A_2|^2 \, \e^{-4 V_1 - 2 \alpha \wt{V}}
+ 2 |B_2|^2 \, \e^{4 V_1 + 2 \alpha \wt{V}}
\, , \\
0 \ &= \ 
- \alpha |A_2|^2 \, \e^{2 V_1 - 2 \alpha \wt{V}}
+ \alpha |B_2|^2 \, \e^{- 2 V_1 + 2 \alpha \wt{V}}
\, .
\end{align}
\esubeq
Here we set $\sqrt{2} \, t^1_1 \equiv + \a^2$.
The equations of motion for $\Phi_1$ and $\wt{\Phi}$ are given as
\bsubeq \label{EOM-Phi-A1}
\begin{align}
0 \ &= \ 
A_1 B_1 - 2 A_2 B_2 + A_3 B_3
\, , \\
0 \ &= \ 
A_2 B_2
\, . 
\end{align}
\esubeq
We find that the charged hypermultiplet $\{ A_2, B_2 \}$ becomes zero.
This implies that they are gauged away by the vector multiplet $\{ \wt{V}, \wt{\Phi} \}$.
The remaining charged multiplets are $\{ A_1, A_3, B_1, B_3 \}$ under the field equations (\ref{EOM-V-A1}) and (\ref{EOM-Phi-A1}).
The solution is given as
\bsubeq \label{A1-A1A3}
\begin{gather}
|B_3| \ = \ |A_1|
\, , \ls
|B_1| \ = \ |A_3|
\, , \\
\e^{2 V_1} 
\ = \
\frac{\a^2 + \sqrt{\a^4 + 4 (|A_1|^2 + |A_3|^2)^2}}{2 (|A_1|^2 + |A_3|^2)}
\, .
\end{gather}
\esubeq
Notice that the set $\{ B_3 , B_1 \}$ becomes the copy of the other one $\{ A_1, A_3 \}$.
Substituting the solution into the Lagrangian (\ref{GLSMF44-A1}) under the IR limit $e_1, \wt{e} \to \infty$,
we obtain 
\begin{align}
\Scr{L}_{\text{IR}}
\ &= \
\int \d^4 \theta \, \Big\{
\sqrt{\a^4 + 4 (|A_1|^2 + |A_3|^2)^2}
- \a^2 \log \Big( \frac{\a^2 + \sqrt{\a^4 + 4 (|A_1|^2 + |A_3|^2)^2}}{2 (|A_1|^2 + |A_3|^2)} \Big)
\Big\} 
\nn \\
\ & \ \ \ \ 
- \sqrt{2} \, t^1_2 \, F_{01}^1  
\, . \label{A1-NLSM}
\end{align}
The first line of the right-hand side describes the K\"{a}hler potential \cite{Gibbons:1979xn} of the Eguchi-Hanson space \cite{Eguchi:1978xp}\footnote{One of the authors have derived the same K\"{a}hler potential in different formulations \cite{Higashijima:2001yn, Higashijima:2001de, Higashijima:2001vk, Higashijima:2001fp}.}.
The parameter $\a$ is the size of a two-sphere which resolves the singularity of the space.
The second line is a topological term which manages the instanton corrections.
If one considers the string worldsheet instanton corrections, i.e., the effect of wrapping string on the two-sphere,
one can analyze the vortex corrections of the original gauge theory (\ref{GLSMF44-A1}).
In the vanishing limit $\a \to 0$, 
the Lagrangian (\ref{A1-NLSM}) becomes
\begin{align}
\Scr{L}_{\text{IR}}^{\text{singular}}
\ &= \ 
2 \int \d^4 \theta \, \Big( |A_1|^2 + |A_3|^2 \Big) 
- \sqrt{2} \, t^1_2 \, F_{01}^1  
\, . \label{A1-NLSM-singular}
\end{align}
This effective theory seems to be a free theory on ${\mathbb C}^2$.
However, 
because the hypermultiplet $\{ B_3, B_1 \}$ is identical to $\{ A_1, A_3 \}$ in (\ref{A1-A1A3}),
the genuine target space is ${\mathbb C}^2/{\mathbb Z}_2$.
This is nothing but the singular limit of the $A_1$-type ALE space.

Let us further analyze the sigma model (\ref{A1-NLSM}).
We parametrize the scalar component fields of the chiral superfields in the following forms:
\bsubeq \label{EH-Eulercoord}
\begin{alignat}{2}
A_1 \ &= \ 
\frac{r}{\sqrt{2}} \cos \frac{\vartheta}{2} \, \e^{\frac{\i}{2} (\psi + \varphi)}
\, , &\ls
A_3 \ &= \ 
\frac{r}{\sqrt{2}} \sin \frac{\vartheta}{2} \, \e^{\frac{\i}{2} (\psi - \varphi)}
\, , \\
B_1 \ &= \ 
\frac{r}{\sqrt{2}} \sin \frac{\vartheta}{2} \, \e^{\frac{\i}{2} (\psi - \varphi)}
\, , &\ls
B_3 \ &= \ 
- \frac{r}{\sqrt{2}} \cos \frac{\vartheta}{2} \, \e^{\frac{\i}{2} (\psi + \varphi)}
\, ,
\end{alignat}
\esubeq
where $\vartheta \in [0, \pi]$, 
$\varphi \in [0 , 2 \pi]$, and
$\psi \in [0, 4 \pi]$ are the Euler angles and $r$ is the radial coordinate of the four-dimensional space.
The ${\mathbb Z}_2$ discrete symmetry of the solution (\ref{A1-A1A3}) generates the ${\mathbb Z}_2$ orbifolding of the Euler angle $\psi$.
Plugging them into the Lagrangian (\ref{A1-NLSM}), 
we obtain 
\begin{align}
\Scr{L}_{\text{IR}}
\ &= \ 
- \Big( 1 - \frac{\a^4}{\rho^4} \Big)^{-1} (\del_m \rho)^2
- \Big( 1 - \frac{\a^4}{\rho^4} \Big)
\frac{\rho^2}{4} \Big\{ (\del_m \psi) + (\del_m \varphi) \cos \vartheta \Big\}^2
- \frac{\rho^2}{4}
\Big\{ (\del_m \vartheta)^2 + (\del_m \varphi)^2 \sin^2 \vartheta \Big\}
\nn \\
\ & \ \ \ \ 
- \sqrt{2} \, t^1_2 \, F_{01}^1
+ \text{(fermionic parts)}
\, . \label{A1-NLSM-2}
\end{align}
Here we redefined $\rho^2 = \sqrt{\a^4 + r^4}$.
This is the NLSM whose target space is represented in terms of the Eguchi-Hanson metric in the well-known form \cite{Eguchi:1978xp}.
In the vanishing limit $\a \to \infty$, 
this is reduced to the sigma model for ${\mathbb C}^2/{\mathbb Z}_2$, explicitly.

In the end of this section,
we emphasize that the existence of F-term is inevitable to find the correct form of the K\"{a}hler potential and the discrete symmetry.
We learned that the D-term associated with the toric data is not sufficient to describe the toric varieties in the viewpoint of differential geometry.

\section{$A_N$-type ALE space}
\label{GLSMF-AN}

In this section we investigate the $\N=(4,4)$ GLSM for a generic $A_N$-type ALE space 
where $N = 2m-1$ (odd) or $N = 2m$ (even).
Following the toric data \cite{Hori:2003ic},
we can immediately extend it to the one for the $\N=(4,4)$ system.
First, we exhibit the field contents for the $\N=(4,4)$ GLSM.
Next, we analyze the IR limit.
Integrating out all the vector multiplets and
taking the singular limit where all the FI parameters vanish,
we obtain the NLSM for the orbifold limit ${\mathbb C}^2/{\mathbb Z}_{N+1}$.
 
Let us prepare $N+2$ charged hypermultiplets $\{ A_i, B_i \}$
and $N+1$ vector multiplets $\{ V_a, \Phi_a; \wt{V}, \wt{\Phi} \}$.
The $U(1)$ charge assignments for the charged hypermultiplets are summarized in Table \ref{table:AN}:
\begin{table}[h]
\begin{center}
\slb{.9}{
$\renewcommand{\arraystretch}{1.05}
\begin{array}{c|ccccccc}
\text{$A_N$-type} & (A_1, B_1) & (A_2, B_2) & \cdots & (A_{m+1}, B_{m+1}) & \cdots & (A_{N+1}, B_{N+1}) & (A_{N+2}, B_{N+2}) 
\\ \hline\hline
(V_1, \Phi_1) & (+1,-1) & (-2,+2) & \cdots & 0 & 0 & 0 & 0
\\ 
(V_2, \Phi_2) & 0 & (+1,-1) & \cdots & \cdots & 0 & 0 & 0
\\ 
\vdots
\\
(V_{m}, \Phi_{m}) & 0 & 0 & (+1,-1) & (-2,+2) & (+1,-1) & 0 & 0
\\ 
\vdots
\\ 
(V_{N}, \Phi_{N}) & 0 & 0 & 0 & 0 & \cdots & (-2,+2) & (+1,-1) 
\\ 
(\wt{V}, \wt{\Phi}) & 0 & 0 & 0 & (- \alpha, \alpha) & 0 & 0 & 0
\end{array}
$
} 
\end{center}
\caption{Field contents in $\N=(4,4)$ GLSM for $A_N$-type ALE space.}
\label{table:AN}
\end{table}

Due to the $\N=(4,4)$ supersymmetry, the Lagrangian is uniquely determined as follows:
\bsubeq \label{GLSMF44-AN}
\begin{align}
\Scr{L}
\ &= \
\int \d^4 \theta \, \Big\{
\sum_{a=1}^{N} \frac{1}{e_a^2} \Big( - |\Sigma_a|^2 + |\Phi_a|^2 \Big)
+ \frac{1}{\wt{e}^2} \Big( - |\wt{\Sigma}|^2 + |\wt{\Phi}|^2 \Big)
\Big\}
\nn \\
\ & \ \ \ \ 
+ \int \d^4 \theta \, \sum_{k=1}^{m} \Big\{
|A_k|^2 \, \e^{2 V_{k-2} - 4 V_{k-1} + 2 V_{k}}
+ |B_k|^2 \, \e^{- 2 V_{k-2} + 4 V_{k-1} - 2 V_{k}}
\Big\}
\nn \\
\ & \ \ \ \ 
+ \int \d^4 \theta \, \Big\{
|A_{m+1}|^2 \, \e^{2 V_{m-1} - 4 V_{m} + 2 V_{m+1} - 2 \alpha \wt{V}}
+ |B_{m+1}|^2 \, \e^{- 2 V_{m-1} + 4 V_{m} - 2 V_{m+1} + 2 \alpha \wt{V}}
\Big\}
\nn \\
\ & \ \ \ \ 
+ \int \d^4 \theta \, \sum_{\l={m+2}}^{N+2} \Big\{
|A_\l|^2 \, \e^{2 V_{\l-2} - 4 V_{\l-1} + 2 V_{\l}}
+ |B_\l|^2 \, \e^{- 2 V_{\l-2} + 4 V_{\l-1} - 2 V_{\l}}
\Big\}
\nn \\
\ & \ \ \ \ 
+ \Big\{
\sqrt{2} \int \d^2 \theta \, \Big[
\sum_{a=1}^{N} \Phi_a \Big( - A_a B_a + 2 A_{a+1} B_{a+1} - A_{a+2} B_{a+2}  \Big)
+ \wt{\Phi} \Big( \alpha A_{m+1} B_{m+1} \Big)
+ \text{(h.c.)}
\Big]
\Big\}
\nn \\
\ & \ \ \ \ 
+ \Big\{
\sqrt{2} \int \d^2 \wt{\theta} \, \sum_{a=1}^{N} \big( - t^a \Sigma_a \big)
+ \text{(h.c.)}
\Big\}
\, . 
\end{align}
For a simple description, 
we introduce dummy variables $\{ V_{-1}, V_0, V_{N+1}, V_{N+2} \}$ and adopt the following conventions:
\begin{align}
\e^{2 V_{-1}} \ &= \ 1 \ = \ \e^{2 V_0}
\, , \ls
\e^{2 V_{N+1}} \ = \ 1 \ = \ \e^{2 V_{N+2}}
\, .
\end{align}
\esubeq

Consider the low energy physics of (\ref{GLSMF44-AN}) in the IR limit $e_a, \wt{e} \to \infty$.
In this limit all the vector multiplets become auxiliary fields
because their kinetic terms disappear.
In order to integrate them out from the system,
we evaluate the field equations.
The field equations for $\{ \wt{V}, V_a \}$ are
\bsubeq \label{EOM-V-AN}
\begin{align}
0 \ &= \ 
- \alpha |A_{m+1}|^2 \, \e^{2 V_{m-1} - 4 V_m + 2 V_{m+1} - 2 \alpha \wt{V}}
+ \alpha |B_{m+1}|^2 \, \e^{- 2 V_{m-1} + 4 V_m - 2 V_{m+1} + 2 \alpha \wt{V}}
\, , \\
\sqrt{2} t^k_1
\ &= \ 
|A_k|^2 \, \e^{2 V_{k-2} - 4 V_{k-1} + 2 V_k}
- |B_k|^2 \, \e^{- 2 V_{k-2} + 4 V_{k-1} - 2 V_k}
\nn \\
\ & \ \ \ \ 
- 2 |A_{k+1}|^2 \, \e^{2 V_{k-1} - 4 V_k + 2 V_{k+1}}
+ 2 |B_{k+1}|^2 \, \e^{- 2 V_{k-1} + 4 V_k - 2 V_{k+1}}
\nn \\
\ & \ \ \ \ 
+ |A_{k+2}|^2 \, \e^{2 V_{k} - 4 V_{k+1} + 2 V_{k+2}}
- |B_{k+2}|^2 \, \e^{- 2 V_{k} + 4 V_{k+1} - 2 V_{k+2}}
\, , \\
\sqrt{2} t^{\l}_1
\ &= \ 
|A_\l|^2 \, \e^{2 V_{\l-2} - 4 V_{\l-1} + 2 V_\l}
- |B_\l|^2 \, \e^{- 2 V_{\l-2} + 4 V_{\l-1} - 2 V_\l}
\nn \\
\ & \ \ \ \ 
- 2 |A_{\l+1}|^2 \, \e^{2 V_{\l-1} - 4 V_\l + 2 V_{\l+1}}
+ 2 |B_{\l+1}|^2 \, \e^{- 2 V_{\l-1} + 4 V_\l - 2 V_{\l+1}}
\nn \\
\ & \ \ \ \ 
+ |A_{\l+2}|^2 \, \e^{2 V_{\l} - 4 V_{\l+1} + 2 V_{\l+2}}
- |B_{\l+2}|^2 \, \e^{- 2 V_{\l} + 4 V_{\l+1} - 2 V_{\l+2}}
\, ,
\end{align}
\esubeq
where $1 \leq k \leq m$, $m+2 \leq \l \leq N+2$.
The field equations for $\{ \wt{\Phi}, \Phi_a \}$ are
\bsubeq \label{EOM-Phi-AN}
\begin{alignat}{2}
0 \ &= \ 
A_{m+1} B_{m+1}
\, , & \ \ \
0 \ &= \ 
A_m B_m - 2 A_{m+1} B_{m+1} + A_{m+2} B_{m+2}
\, , \\
0 \ &= \ 
A_{m-1} B_{m-1} - 2 A_m B_m + A_{m+1} B_{m+1}
\, , & \ \ \ 
0 \ &= \ 
A_{m+1} B_{m+1} - 2 A_{m+2} B_{m+2} + A_{m+3} B_{m+3}
\, , \\
&\LS \vdots
& \ \ \ 
&\LS \vdots 
\nn \\
0 \ &= \ 
A_1 B_1 - 2 A_2 B_2 + A_3 B_3
\, , & \ \ \ 
0 \ &= \ 
A_{N} B_{N} - 2 A_{N+1} B_{N+1} + A_{N+2} B_{N+2}
\, . 
\end{alignat}
\esubeq
The field equations for $\{ \wt{V}, \wt{\Phi} \}$ require that 
the charged hypermultiplet $\{ A_{m+1} , B_{m+1} \}$ vanishes.
Unfortunately, however, 
it is difficult to solve the equations (\ref{EOM-V-AN}) 
if the FI parameters $t^a_1$ are non-zero.
Thus we focus only on the case where all the FI parameters vanish.
Furthermore, in order to simplify the equations (\ref{EOM-Phi-AN}),
we recombine the equations (\ref{EOM-Phi-AN}) with each other as in the following forms:
\bsubeq \label{EOM-Phi-AN-2}
\begin{alignat}{2}
0 \ &= \ 
A_{m+1} \ = \ B_{m+1}
\, , &\ls
0 \ &= \ 
A_m B_m + A_{m+2} B_{m+2}
\, , \\
0 \ &= \ 
A_{m-1} B_{m-1} + 2 A_{m+2} B_{m+2}
\, , &\ls
0 \ &= \ 
A_{m+3} B_{m+3} + 2 A_m B_m
\, , \\
0 \ &= \ 
A_{m-2} B_{m-2} + 3 A_{m+2} B_{m+2}
\, , &\ls
0 \ &= \ 
A_{m+4} B_{m+4} + 3 A_m B_m
\, , \\
&\LS \vdots
&\ls
&\LS \vdots
\nn \\
0 \ &= \ 
A_k B_k + (m+1-k) A_{m+2} B_{m+2}
\, , &\ls
0 \ &= \ 
A_{\l} B_{\l} + (\l - m-1) A_m B_m
\, , \\
&\LS \vdots
&\ls
&\LS \vdots
\nn \\
0 \ &= \ 
A_1 B_1 + m \, A_{m+2} B_{m+2}
\, , &\ls
0 \ &= \ 
A_{N+2} B_{N+2} + m' A_m B_m
\, , 
\end{alignat}
\esubeq
where $m' = m$ (if $N=2m-1$) or $m'=m+1$ (if $N=2m$).
We can analytically solve the field equations (\ref{EOM-V-AN}) and (\ref{EOM-Phi-AN-2}).
The solution is 
\bsubeq \label{AN-A1AN}
\begin{alignat}{3}
|A_k| \ &= \ 
\sqrt{\frac{m+1-k}{m}} \, |A_1|
\, , &\ls
|A_{m+1}| \ &= \ 0 
\, , &\ls
|A_\l| \ &= \ 
\sqrt{\frac{\l - (m+1)}{m'}} \, |A_{N+2}|
\, , \\
|B_k| \ &= \ 
\sqrt{\frac{m+1-k}{m'}} \, |A_{N+2}|
\, , &\ls
|B_{m+1}| \ &= \ 0 
\, , &\ls
|B_\l| \ &= \ 
\sqrt{\frac{\l - (m+1)}{m}} \, |A_1|
\, , \\
&&
\e^{2 V_a} \ &= \ 1
\, .
\end{alignat}
\esubeq
The solution has 
one dynamical hypermultiplet $\{ A_1 , A_{N+2} \}$ and its $N$ copies $\{ A_i, B_i \}$ up to coefficients.
There exists a ${\mathbb Z}_{N+1}$ symmetry under the rotation among the $N+1$ pairs of the hypermultiplets.
Plugging this into the Lagrangian (\ref{GLSMF44-AN})
under the IR limit $e_a, \wt{e} \to \infty$, 
we obtain
\begin{align}
\Scr{L}_{\text{IR}}^{\text{singular}}
\ &= \ 
(m+1) \int \d^4 \theta \, \Big( \frac{m'}{m} |A_1|^2 + |A_{N+2}|^2 \Big)
- \sum_{a=1}^{N} \sqrt{2} \, t^a_2 \, F_{01}^a
\, . \label{AN-NLSM-singular}
\end{align}
This is the NLSM for the singular limit ${\mathbb C}^2/{\mathbb Z}_{N+1}$ of the $A_N$-type ALE space.
The rotational symmetry of the solution (\ref{AN-A1AN}) in the gauge theory is the origin of the ${\mathbb Z}_{N+1}$ orbifold symmetry of the geometry.
If we adopt a different arrangement of (\ref{EOM-Phi-AN}) from (\ref{EOM-Phi-AN-2}),
we find a different solution.
The arrangement of the equations (\ref{EOM-Phi-AN}) could be interpreted as the coordinate transformations on the target space of the NLSM.
We understand that the existence of the F-term is inevitable
to derive the correct NLSM (\ref{AN-NLSM-singular}),
as discussed in the case of the $A_1$-type ALE space.
In the case of the $A_2$-type ALE space (see appendix \ref{GLSMF-A2}), 
we successfully obtained the K\"{a}hler potential involving finite values of the FI parameters. 

Finally, let us argue the role of the topological terms in (\ref{AN-NLSM-singular}).
As in the case of the $A_1$-type ALE space,
the instanton corrections to the NLSM in the IR regime 
can be traced by the vortex corrections of the gauge theory in the UV regime.
They are governed by $N$ independent gauge fields $F_{01}^a$.
Each gauge field makes a vortex configuration, and
each vortex deforms the parameter $t^a_2$.
Since each gauge sector is completely independent of one another,
we can control the deformation of each singularity point independently.
In the viewpoint of the target space,
the deformation of $t^a_2$ implies the wrapping of string around the corresponding singularity point.
Then we conclude that the GLSM (\ref{GLSMF44-AN}) governs the stringy corrections of the $A_N$-type ALE space in a simple way.

\section{Summary and discussions}
\label{Summary}

In this paper 
we constructed the $\N=(4,4)$ GLSMs for A-type ALE spaces
in which the toric descriptions are manifest.
One of the crucial developments is that 
we systematically introduced the F-term which governs the complex structure of the target space geometry.
The existence of the F-term is inevitable 
when we explicitly analyze the NLSM for the toric varieties by virtue of the metric, curvature, and other objects of differential geometry.

First, we exhibited the toric data of the $A_1$-type ALE space, i.e., the Eguchi-Hanson space.
We applied the toric data to the $\N=(4,4)$ system by introducing additional supermultiplets and the $SU(2)_R$ symmetry.
The extended supersymmetry naturally generates the F-term.
In the IR limit, 
we explicitly derived the K\"{a}hler potential of the Eguchi-Hanson space.
Furthermore, we also found the origin of the ${\mathbb Z}_2$ orbifold symmetry as the discrete rotational symmetry in the solution of the gauge theory.

Next, we applied the same technique to the $\N=(4,4)$ GLSM for the $A_N$-type ALE space.
Since the construction rule is highly systematic,
the Lagrangian is uniquely determined 
once we prepare the field contents following the toric data.
The F-term is also automatically provided by the $SU(2)_R$ symmetry.
We successfully obtained the K\"{a}hler potential of the singular limit of the $A_N$-type ALE space,
although it is difficult to solve the equations of motion in the presence of the FI parameters.
In a specific case $N = 2$, we obtained the K\"{a}hler potential involving the finite values of the FI parameters (see appendix \ref{GLSMF-A2}).

The $\N=(4,4)$ GLSM (\ref{GLSMF44-AN}) provides non-trivial topological terms governed by the gauge fields $F_{01}^a$ in the NLSM (\ref{AN-NLSM-singular}).
They are associated with the singularity points on the ALE space.
When we study the worldsheet instanton corrections by the wrapping string along the singularity points,
we analyze the vortex corrections in the gauge theory regime.
Since each sector of the gauge symmetries is completely independent of one another,
we can control the deformations of singularities points-by-points.

\vspace{3mm}

In order to remove redundant degrees of freedom in the gauge theory,  
we introduced the vector multiplet $\{ \wt{V}, \wt{\Phi} \}$.
This multiplet is not associated with the toric data in our construction.
It will be interesting if we import this gauge multiplet into the geometrical feature of toric variety.

The $A_N$-type ALE space is the transverse space of $N+1$ parallel Kaluza-Klein (KK) five-branes.
The $\N=(4,4)$ GLSM for such a system is also suggested in \cite{Tong:2002rq, Harvey:2005ab, Okuyama:2005gx}, 
where the neutral hypermultiplet becomes the coordinate fields in the IR limit.
This is different from the GLSM (\ref{GLSMF44-AN}).
Indeed, there are various GLSMs to describe the $A_N$-type ALE space 
(see, for instance, \cite{Douglas:1996sw, Johnson:1996py, Giveon:1998sr, Belhaj:2000ai}).
Our GLSM will provide a new approach to investigate various five-branes such as KK five-branes, NS5-branes \cite{Tong:2002rq},
and an exotic five-brane \cite{Kimura:2013fda, Kimura:2013zva, Kimura:2013khz}.

One of the most ambitious works is to apply a similar technique to the $\N=(2,2)$ GLSM for toric Calabi-Yau varieties.
A typical example is the singular (or resolved) conifold.
This geometry has been utilized widely in topological string.
Unfortunately, however, 
it is difficult to derive the correct metric of the conifold 
only in terms of the toric data \cite{Oota:2005mr}.
The same difficulty also appears in the $\N=(2,2)$ toric GLSM for the conifold.
Because the conifold is a K\"{a}hler space rather than a hyper-K\"{a}hler,
it is impossible to introduce an $SU(2)_R$ symmetry associated with the $\N=(4,4)$ supersymmetry.
Instead of the $SU(2)_R$ symmetry,
we have to find a suitable rule how to introduce the correct F-term into the $\N=(2,2)$ GLSM which governs the complex structure of the conifold.
Furthermore, 
even if we find the correct F-term in the $\N=(2,2)$ GLSM,
we have to check occurrence of deformations of the target space geometry caused by the renormalization group (RG) flow.
Indeed, we should solve the Monge-Amp\`{e}re equation to understand the deformations by the RG flow, in cases of generic Calabi-Yau varieties.
It is hard to analyze the Monge-Amp\`{e}re equation because this is a complicated partial differential equation.
Fortunately, in the case of the conifold,
it is reported that we need not investigate the Monge-Amp\`{e}re equation \cite{Oota:2005mr}.
This implies that it would be enough to focus on the construction of the correct F-term in the $\N=(2,2)$ GLSM in the classical level.

\section*{Acknowledgements}

The authors thank
Shogo Aoyama,
Tohru Eguchi,
Shun'ya Mizoguchi,
Kazutoshi Ohta,
Kazumi Okuyama,
Makoto Sakaguchi,
Shin Sasaki,
Yukinori Yasui
and 
Yutaka Yoshida
for discussions and comments.
The work of TK is supported in part by the MEXT-Supported Program for the Strategic Research Foundation at Private Universities 
from MEXT (Ministry of Education, Culture, Sport, Science and Technology) of Japan (S0901029: Research Center for Measurement in Advanced Science, Rikkyo University).
The work of MY is supported in part by the JSPS Research Fellowship for Young Scientists (\#24-05404).

\begin{appendix}
\section*{Appendix}

\section{$A_2$-type ALE space}
\label{GLSMF-A2}

In this appendix we discuss the $\N=(4,4)$ GLSM for $A_2$-type ALE space.
This is another example that we can obtain the K\"{a}hler potential of the geometry on which the singularities are blown-up.
Following the toric data given in \cite{Hori:2003ic},
we prepare the field contents suitable to the $\N=(4,4)$ supersymmetry as in Table \ref{table:A2}:
\begin{table}[h]
\begin{center}
\slb{.9}{${\renewcommand{\arraystretch}{1.05}
\begin{array}{c|cccc}
\text{$A_{2}$-type} & (A_1, B_1) & (A_2, B_2) & (A_3, B_3) & (A_4, B_4) 
\\ \hline\hline
(V_1, \Phi_1) & (+1,-1) & (-2,+2) & (+1,-1) & 0 
\\ 
(V_2, \Phi_2) & 0 & (+1,-1) & (-2,+2) & (+1,-1) 
\\ 
(\wt{V}, \wt{\Phi}) & 0 & (-\alpha, \alpha) & 0 & 0
\end{array}
}$} 
\end{center}
\caption{Field contents in $\N=(4,4)$ GLSM for $A_2$-type ALE space.}
\label{table:A2}
\end{table}

Due to the $\N=(4,4)$ supersymmetry, the Lagrangian is uniquely determined as
\begin{align}
\Scr{L}
\ &= \
\int \d^4 \theta \, \Big\{
\frac{1}{e_1^2} \Big( - |\Sigma_1|^2 + |\Phi_1|^2 \Big)
+ \frac{1}{e_2^2} \Big( - |\Sigma_2|^2 + |\Phi_2|^2 \Big)
+ \frac{1}{\wt{e}^2} \Big( - |\wt{\Sigma}|^2 + |\wt{\Phi}|^2 \Big)
\Big\}
\nn \\
\ & \ \ \ \ 
+ \int \d^4 \theta \, \Big\{
|A_1|^2 \, \e^{2 V_1}
+ |A_2|^2 \, \e^{-4 V_1 + 2 V_2 - 2 \alpha \wt{V}}
+ |A_3|^2 \, \e^{2 V_1 - 4 V_2}
+ |A_4|^2 \, \e^{2 V_2}
\Big\}
\nn \\
\ & \ \ \ \ 
+ \int \d^4 \theta \, \Big\{
|B_1|^2 \, \e^{-2 V_1}
+ |B_2|^2 \, \e^{4 V_1 - 2 V_2 + 2 \alpha \wt{V}}
+ |B_3|^2 \, \e^{-2 V_1 + 4 V_2}
+ |B_4|^2 \, \e^{-2 V_2}
\Big\}
\nn \\
\ & \ \ \ \ 
+ \Big\{
\sqrt{2} \int \d^2 \theta \, \Big[
\Phi_1 \Big( - A_1 B_1 + 2 A_2 B_2 - A_3 B_3 \Big)
+ \Phi_2 \Big( - A_2 B_2 + 2 A_3 B_3 - A_4 B_4 \Big)
+ \text{(h.c.)}
\Big]
\Big\}
\nn \\
\ & \ \ \ \ 
+ \Big\{
\sqrt{2} \int \d^2 \theta \, 
\wt{\Phi} \Big( \alpha A_2 B_2 \Big)
+ \text{(h.c.)}
\Big\}
\nn \\
\ & \ \ \ \ 
+ \Big\{
\sqrt{2} \int \d^2 \wt{\theta} \, \big( - t^1 \Sigma_1 - t^2 \Sigma_2 \big)
+ \text{(h.c.)}
\Big\}
\, . \label{GLSMF44-A2}
\end{align}
Here $e_a$ and $\wt{e}$ are the gauge coupling constants of mass dimension one.
We also introduced the complexified FI parameters $t^1$ and $t^2$ associated with gauge multiplets $\Sigma_1 = \frac{1}{\sqrt{2}} \ol{D}{}_+ D_- V_1$ and $\Sigma_2 = \frac{1}{\sqrt{2}} \ol{D}{}_+ D_- V_2$, respectively.
However, we do not introduce the FI parameters associated with 
$\Phi_1$, $\Phi_2$, 
$\wt{\Sigma} = \frac{1}{\sqrt{2}} \ol{D}{}_+ D_- \wt{V}$, and $\wt{\Phi}$.

We investigate the dynamics in the IR limit $e_a, \wt{e} \to \infty$,
where all the $\N=(4,4)$ vector multiplets become auxiliary fields.
Then we completely integrate them out from the system.
The field equations for $\{ V_a, \wt{V} \}$ are
\bsubeq \label{EOM-V-A2}
\begin{align}
\sqrt{2} \, t^1_1
\ &= \ 
|A_1|^2 \, \e^{2 V_1} 
- |B_1|^2 \, \e^{-2 V_1}
- 2 |A_2|^2 \, \e^{-4 V_1 + 2 V_2 - 2 \alpha \wt{V}}
+ 2 |B_2|^2 \, \e^{4 V_1 - 2 V_2 + 2 \alpha \wt{V}}
\nn \\
\ & \ \ \ \ 
+ |A_3|^2 \, \e^{2 V_1 - 4 V_2}
- |B_3|^2 \, \e^{- 2 V_1 + 4 V_2}
\, , \\
\sqrt{2} \, t^2_1
\ &= \ 
|A_2|^2 \, \e^{-4 V_1 + 2 V_2 - 2 \alpha \wt{V}}
- |B_2|^2 \, \e^{4 V_1 - 2 V_2 + 2 \alpha \wt{V}}
- 2 |A_3|^2 \, \e^{2 V_1 - 4 V_2}
+ 2 |B_3|^2 \, \e^{- 2 V_1 + 4 V_2}
\nn \\
\ & \ \ \ \ 
+ |A_4|^2 \, \e^{2 V_2}
- |B_4|^2 \, \e^{-2 V_2}
\, , \\
0 \ &= \ 
- \alpha |A_2|^2 \, \e^{- 4 V_1 + 2 V_2 - 2 \alpha \wt{V}}
+ \alpha |B_2|^2 \, \e^{4 V_1 - 2 V_2 + 2 \alpha \wt{V}}
\, .
\end{align}
\esubeq
The field equations for $\{ \Phi_a, \wt{\Phi} \}$ are 
\bsubeq \label{EOM-Phi-A2}
\begin{align}
0 \ &= \ 
A_1 B_1 - 2 A_2 B_2 + A_3 B_3
\, , \\
0 \ &= \ 
A_2 B_2 - 2 A_3 B_3 + A_4 B_4
\, , \\
0 \ &= \ 
A_2 B_2
\, . 
\end{align}
\esubeq
The charged hypermultiplet $\{ A_2, B_2 \}$ is gauged away by 
the field equations for $\{ \wt{V}, \wt{\Phi} \}$. 
Here we introduce a relation between the two FI parameters 
\begin{align}
\sqrt{2} \, t^1_1 \ &\equiv \ \pm \a^2 \ \equiv \ - \sqrt{2} \, t^2_1
\, . \label{FI-A2}
\end{align}
This implies that the two singularities on the space are blown up by 
two ${\mathbb C}{\bf P}^1$ with the same size, except for the difference of the relative signs.
Only in this case we find the analytical solution of the equations (\ref{EOM-V-A2}) and (\ref{EOM-Phi-A2}):
\bsubeq \label{A2-A1A4}
\begin{gather}
\begin{alignat}{4}
|A_1| &
\, , &\ \ \ 
|A_2| \ &= \ 0
\, , &\ \ \ 
|A_3| \ &= \ \frac{1}{\sqrt{2}} |B_4|
\, , &\ \ \ 
|A_4| \ &= \ \sqrt{2} |A_1|
\, , \\
|B_1| \ &= \ \frac{1}{\sqrt{2}} |B_4|
\, , &\ \ \ 
|B_2| \ &= \ 0
\, , &\ \ \ 
|B_3| \ &= \ |A_1|
\, , &\ \ \ 
|B_4| \ &
\, ,
\end{alignat}
\\
\e^{2 V_1} 
\ = \
\frac{\pm \a^2 + \sqrt{\a^4 + 8 |A_1 B_4|^2}}{4 |A_1|^2}
\, , \ls
\e^{2 V_2}
\ = \ 
\frac{|B_4|}{\sqrt{2}|A_1|}
\, . 
\end{gather}
\esubeq
In this solution,
there exists a ${\mathbb Z}_3$ rotational symmetry among the three pairs 
$\{ A_1, B_1 \}$, $\{ B_3, A_3 \}$ and $\{ A_4, B_4 \}$ up to coefficients.
Even though there are two analytical solutions caused by the choice of the sign in (\ref{FI-A2}), 
only $\sqrt{2} \, t^1_1 = + \a^2 = - \sqrt{2} \, t^2_1$ is applicable.
The reason is that the dynamical chiral superfield $A_1$ represents the coordinate of the ${\cal O}(-2)$ bundle over the first ${\mathbb C}{\bf P}^1$ if we set $t^1_1$ to be positive, in the framework of the toric description \cite{Witten:1993yc, Hori:2003ic}.
The same interpretation is also applicable from the viewpoint of the dynamical chiral superfield $B_4$.
In order that $B_4$ represents the coordinate of the ${\cal O}(-2)$ bundle over the second ${\mathbb C}{\bf P}^1$, 
we have to set $t^2_1$ to be negative.
If we choose the negative sign in (\ref{FI-A2}),
the above geometrical interpretation does not make sense.
Plugging (\ref{A2-A1A4}) into the GLSM under the IR limit $e_a, \wt{e} \to \infty$, we obtain 
\begin{align}
\Scr{L}_{\text{IR}}
\ &= \ 
\int \d^4 \theta \, \Big\{
2 \sqrt{2} |A_1 B_4|
+ \sqrt{\a^4 + 8 |A_1 B_4|^2} 
- \a^2 \log \Big( \frac{\a^2 + \sqrt{\a^4 + 8 |A_1 B_4|^2}}{2 \sqrt{2} |A_1 B_4|} \Big) 
\Big\}
\nn \\
\ & \ \ \ \ 
- \sqrt{2} \, t^1_2 \, F_{01}^1  
- \sqrt{2} \, t^2_2 \, F_{01}^2 
\, . \label{A2-NLSM}
\end{align}
This is the NLSM for the $A_2$-type ALE space, 
where each singularity is blown up by ${\mathbb C}{\bf P}^1$ of size $\a$.
There are two topological terms in the second line of the right-hand side.
Since they are governed by two different gauge fields independently,
we can argue the worldsheet instanton corrections to the two two-spheres separately, even though the sizes of the two ${\mathbb C}{\bf P}^1$ are same.

Finally,
we consider the singular limit $\a \to 0$.
The Lagrangian (\ref{A2-NLSM}) is reduced to
\begin{align}
\Scr{L}_{\text{IR}}
\ &= \ 
4 \sqrt{2} \int \d^4 \theta \, |A_1 B_4| 
- \sqrt{2} \, t^1_2 \, F_{01}^1  
- \sqrt{2} \, t^2_2 \, F_{01}^2 
\, . \label{A2-NLSM-singular}
\end{align}
This is a bit different from the result (\ref{AN-NLSM-singular}) in section \ref{GLSMF-AN}:
\begin{align}
\Scr{L}_{\text{IR}}
\ &= \ 
2 \int \d^4 \theta \, \Big( 2 |A_1|^2 + |A_{4}|^2 \Big)
- \sqrt{2} \, t^1_2 \, F_{01}^1  
- \sqrt{2} \, t^2_2 \, F_{01}^2 
\, . \label{A2-NLSM-singular-2}
\end{align}
We think that the arrangement of the field equations for $\Phi_a$ (\ref{EOM-Phi-AN-2}) in the gauge theory
can be interpreted as a coordinate transformation (and a change of dynamical fields) from the one in (\ref{A2-NLSM-singular}) to the other in (\ref{A2-NLSM-singular-2}) in the IR limit.
We mention that both of the models (\ref{A2-NLSM-singular}) and (\ref{A2-NLSM-singular-2}) represent the NLSMs for ${\mathbb C}^2/{\mathbb Z}_3$ orbifold.

\end{appendix}

}
\end{document}